# A Risk-Averse Just-In-Time Scheme for Learning-Based Operation of Microgrids with Coupled Electricity-Hydrogen-Ammonia under Uncertainties

Longyan Li, *Student Member, IEEE*, Chao Ning, *Senior Member, IEEE*, Guangsheng Pan, *Member, IEEE*, Leiqi Zhang, Wei Gu, *Senior Member, IEEE*, Liang Zhao, Wenli Du, *Senior Member, IEEE*, and Mohammad Shahidehpour, *Life Fellow, IEEE*

*Abstract*—This paper proposes a Risk-Averse Just-In-Time (RAJIT) operation scheme for Ammonia-Hydrogen-based Micro-Grids (AHMGs) to boost electricity-hydrogen-ammonia coupling under uncertainties. First, an off-grid AHMG model is developed, featuring a novel multi-mode ammonia synthesis process and a hydrogen-ammonia dual gas turbine with tunable feed-in ratios. Subsequently, a state-behavior mapping strategy linking hydrogen storage levels with the operation modes of ammonia synthesis is established to prevent cost-ineffective shutdowns. The proposed model substantially improves operational flexibility but results in a challenging nonlinear fractional program. Based upon this model, a data-driven RAJIT scheme is developed for the real-time rolling optimization of AHMGs. Unlike conventional one-size-fits-all schemes using one optimization method throughout, the data-driven RAJIT intelligently switches between cost-effective deterministic optimization and risk-averse online-learning distributionally robust optimization depending on actual risk profiles, thus capitalizing on the respective strengths of these two optimization methods. To facilitate the solution of the resulting nonlinear program, we develop an equivalent-reformulation-based solution methodology by leveraging a constraint-tightening technique. Numerical simulations demonstrate that the proposed scheme guarantees safety and yields an overall cost reduction up to 14.6% compared with several state-of-the-art methods.

## NOMENCLATURE

### A. Sets and Indices

| | |
|---|---|
| $i$ | Index of operation modes. |
| $\mathbb{B}$ | Binary set {0,1}. |
| $t$ | Index of time. |
| $j, k / S^N$ | Index/set of buses. |
| $(j, k) / S^L$ | Index/set of lines. |
| $w/b/p$ | Index of WT/battery/HAP. |
| $s/S^{WT/BT/HAP}$ | Index/sets of buses connected with WT/battery/HAP. |

### B. Parameters

| | |
|---|---|
| $c^{BT/HS}$ | Unit degradation cost of battery/H$_2$ storage. |
| $c^C$ | Unit curtailment cost. |
| $c^{ess/DGT/elz}$ | Unit operation cost of battery/DGT/electrolyzer. |
| $c^{HB}_{SU/SD/AD/M}$ | Start-up/shut-down/adjust/operation cost of HB process. |
| $C_{elz}$ | Transformation factor of electrolyzer. |
| $\overline{E}^{BT}$ | Capacity of battery. |
| $M^{N_2/H_2/NH_3}$ | Molar mass of N$_2$/H$_2$/NH$_3$. |
| $\overline{m}^{HS/NH_3}$ | Capacity of H$_2$/NH$_3$ storage. |
| $m_t^{H_2D/NH_3D}$ | H$_2$/NH$_3$ demand. |
| $N_p$ | Look-ahead horizon. |
| $\{P/Q\}^{\max}_{t(j,k)}$ | Active/reactive power limit of lines. |
| $P_t^{WT/PV/L}$ | Power of WT/PV/load. |
| $\overline{P}^{BT/HB}$ | Maximum power of battery/HB process. |
| $\overline{P}^{DGT/PSA}$ | Capacity of DGT/PSA. |
| $R / X_{(j,k)}$ | Resistance/reactance of lines. |
| $U_j^{\min/\max}$ | Minimum/maximum voltage value of buses. |
| $U_0$ | Voltage reference value. |
| $\delta^{-/+}$ | Ramping down/up rates. |
| $\underline{\eta}^{BT} / \overline{\eta}^{BT}$ | Minimum/maximum SOC of battery. |
| $\eta^{Com/PSA}$ | Unit power consumption of compressor/PSA. |
| $\eta_b^{ch/dch}$ | Efficiency of battery charging/discharging. |
| $\eta^{DGT/HB}$ | Efficiency of DGT/HB process. |

This work was supported in part by the National Natural Science Foundation of China under Grants 62473256 and 62103264, in part by National Natural Science Foundation of China (Basic Science Center Program: 61988101), and in part by the Open Research Project of the State Key Laboratory of Industrial Control Technology, China (Grant No. ICT2024B71). *(Corresponding author: Chao Ning)*

L. Li and C. Ning are with Department of Automation, Shanghai Jiao Tong University, Shanghai 200240, China, and Key Laboratory of System Control and Information Processing, Ministry of Education of China, Shanghai 200240, China. (e-mail: lilongyan@sjtu.edu.cn; chao.ning@sjtu.edu.cn).

G. Pan and W. Gu are with the School of Electrical Engineering, Southeast University, Nanjing 210096, China (e-mail: pgspan@163.com; wgu@seu.edu.cn).

L. Zhang is with the Zhejiang Key Laboratory of Distributed Generations and Microgrid Technology, State Grid Zhejiang Electric Power Research Institute, Hangzhou 310014, China. (e-mail: y_delta@163.com).

L. Zhao and W. Du are with Key Laboratory of Smart Manufacturing in Energy Chemical Process, Ministry of Education, East China University of Science and Technology (e-mail: lzhao@ecust.edu.cn; wldu@ecust.edu.cn).

Mohammad Shahidehpour is with the Department of Electrical and Computer Engineering, Illinois Institute of Technology, Chicago, IL 60616 USA (e-mail: ms@iit.edu).

| | |
|---|---|
| $\eta^{elz/PSA/HB}$ | Efficiency of electrolyzer /PSA/HB process. |
| $\underline{\eta}^{HB} / \overline{\eta}^{HB}$ | Minimum/maximum operation range of HB process. |
| $\eta^{HS}$ | Constant to divide the $H_2$ storage SOC. |
| $\theta_i$ | Power factor of load. |

C. *Variables*

| | |
|---|---|
| $c_t^{C/op/de}$ | Curtailment/operation/degradation cost. |
| $e_t^{BT}$ | Battery storage level. |
| $LHV_t$ | Lower heating value. |
| $m_t^{H_2GTin/H_2HBin}$ | Mass of $H_2$ input of DGT/HB process. |
| $m_t^{HB}$ | $NH_3$ production mass of HB process. |
| $m_t^{HS/NH_3}$ | Hydrogen/ammonia storage level. |
| $q_t^{H_2in/N_2in}$ | Moles of feed-in $H_2/N_2$ of HB process. |
| $P/Q$ | Active/reactive power. |
| $P_t^{ch/dch}$ | Charging/discharging power of battery. |
| $P_t^C$ | Curtailment power. |
| $P_t^{DGT/HAP/PSA/elz}$ | Power of DGT/HAP/PSA/electrolyzer. |
| $P_t^{HB} / P_t^{HB-i}$ | Total/$i$-$th$ mode operation power of HB process. |
| $U_{tj}$ | Voltage value of buses. |
| $r_t^{HB/DGT}$ | Feed-in ratio of HB/DGT. |
| $v_t$ | Power rate changing indicator of HB. |
| $\varepsilon_{it}^{HB/HS}$ | State indicator of HB process/$H_2$ storage. |
| $\eta_t^{CE}$ | Combustion efficiency. |

## I. INTRODUCTION

*A. Motivation*

THE concept of Ammonia-Hydrogen-based Micro-Grids (AHMGs) has gained considerable attention as a compelling solution for the low-carbon transition of energy and industrial sectors [1], [2], [3]. The off-grid AHMG is the game changer for the scale production of green hydrogen ($H_2$) and green ammonia ($NH_3$), particularly in remote locations endowed with abundant renewable energy resources but lacking access to major electricity grids. Off-grid AHMG has already gained substantial traction, with pioneering projects being lunched in many countries, including Australia [4], China [5], Oman [6], and Demark [7]. In AHMGs, harnessing the coupling effect of electricity, $H_2$, and $NH_3$ is of utmost importance for noticeably improving the AHMG operation performance [8], [9]. However, achieving this coupling safely remains a challenge especially in the context of off-grid configurations, primarily due to the uncertainties introduced by loads and renewable energy sources.

*B. Literature Review*

The operation cost stands as a crucial target of the electricity-hydrogen-ammonia coupling in the off-grid AHMG operation [10], [11]. It largely depends on the Haber-Bosch (HB) process, which serves as the fundamental building block of ammonia synthesis yet inherently possesses limited operational flexibility [12], [13]. The authors of [14] considered the HB process to operate at the rated power only when $H_2$ storage exceeded a predetermined threshold, resulting in significant renewable energy curtailment. Departing from the above literature which assumes that the HB process operates at rated power, an increasing number of studies have incorporated flexible operation of the HB process with various load ranges [15], [16], spanning from 20% to 100%, with a ramping rate of 20% [17], [18], [19]. Considering the actual reactor stages, a three-bed model of HB process was designed for an off-grid AHMG [20], with a hydrogen genset and an ammonia genset incorporated for power generation. In these studies, the HB process power was determined by the mass of feed-in materials, and a $H_2$ to nitrogen ($N_2$) feed-in ratio of the feed-in materials was set to be 3:1. A receding-horizon optimization model was developed to optimally schedule dispatch for an AHMG [17], and the power of the HB process was restricted to changing no more than once every four hours. An AHMG was optimally scheduled in accordance with electricity prices [18], revealing enhanced electricity-hydrogen-ammonia coupling indicated by a cost saving of up to 26.17%. Based on the steady-state operation model of the HB process, the authors of [21] further proposed a dynamic regulation process model, making the AHMG more practically applicable and economically profitable. The aforementioned existing literature only centered on the steady-state operation of the HB process with a fixed feed-in ratio and adopted an over-simplified power consumption efficiency.

In addition to operation costs, the operational safety of off-grid AHMGs under uncertainty is also an important aspect. The imperfect forecast of intermittent renewable resources and loads brings about uncertainty, leading to constraint violations that jeopardize operational safety [22], [23], [24]. Based on the treatment of the uncertainty, the existing methods for the off-grid AHMG operation can be categorized into two distinct groups, namely deterministic methods and uncertainty-aware methods [25], [26]. Deterministic optimization methods are typically cost-effective yet fail to explicitly account for uncertainties [27], [28]. Uncertainty-aware methods include stochastic optimization, robust optimization, and Distributionally Robust Optimization (DRO). An optimal daily operation framework based on stochastic optimization for an AHMG was developed in [29]. In reference [30], a robust design methodology was applied for the AHMG. DRO achieves a less conservative solution than robust optimization, by taking advantage of an ambiguity set comprising all possible probability distributions of uncertainty data [31], [32], [33]. However, these studies are all confined to one specific optimization method over the entire operation horizon, limiting their adaptability and robustness in dynamic real-world scenarios with time-varying risk profiles.

*C. Research Gaps*

The existing literature exhibits a growing interest in the off-grid AHMGs. However, there are several shortcomings. First, prior studies tend to overlook the electricity-hydrogen-ammonia coupling in critical devices when modeling off-grid AHMGs. The adjustment of the HB process feed-in ratio is overlooked, despite the potential influence of highly variable renewable energy sources in off-grid settings [34]. The utilization of Dual-fuel Gas Turbines (DGT) in AHMGs has been neglected, not to mention its underlying coupling effect. The DGT employs a dual fuel of $H_2$ and $NH_3$ to generate electricity, and therefore it enjoys better combustion performance in terms of safety and efficiency compared with single-fuel gas turbines [35]. Moreover, the tunable feed-in ratio of the DGT strengthens the electricity-hydrogen-ammonia coupling from a power-supply perspective. Second, the prevailing literature typically adheres to one single real-time operation method, either deterministic or uncertainty-aware optimization. However, the actual risk profile of the off-grid AHMG changes over time, so being either risk-agnostic or risk-averse throughout the whole operation horizon is unnecessary or even counterproductive. In specific, the deterministic method suffers from safety issues during periods of high-risk profile due to its ignorance of uncertainty. Conversely, uncertainty-aware methods, which account for uncertainty throughout the operation horizon, tend to be over-conservative. The comparison of this paper with other literature is tabulated in TABLE I.

TABLE I. Comparison of This Paper with Previous Literature with AHMGs

| Ref. | Flexible feed-in ratio of DGT+HB process | Uncertainty | Model type |
|---|---|---|---|
| [10] | × | × | LP[1] |
| [14] | × | × | NLP[2] |
| [17], [20] | × | × | MILP[3] |
| [18] | × | × | QP[4] |
| [21] | × | √ (Robust) | MINLP[5] |
| [36] | × | √ (Robust) | DP[6] |
| [37] | × | √ (Robust) | LP |
| This study | √ | √ (RAJIT) | MINLFP[7] |

1: linear programming, 2: nonlinear linear programming, 3: mixed integer linear programming, 4: quadratic programming, 5. mixed-integer nonlinear programming, 6. dynamic programming 7. mixed-integer nonlinear fractional programming.

*D. Contributions and Organization*

To fill these two research gaps, this paper proposes an RAJIT real-time operation scheme for off-grid AHMGs in the presence of fluctuant renewable power generation and loads. First, we present an AHMG model including a DGT and a multi-mode HB process, for which a state-behavior mapping strategy is developed linking $H_2$ storage levels with operation modes of the HB process. This model ensures the consistent operation of HB processes even during periods of hydrogen scarcity, and strengthens the electricity-hydrogen-ammonia coupling. Second, a data-driven RAJIT scheme is proposed for the real-time operation of off-grid AHMGs. The AHMG operational status is categorized as safe or unsafe according to its actual risk profile. Then, the data-driven RAJIT scheme switches between a deterministic optimization model and an online-learning distributionally robust optimization model accordingly, thereby being risk-averse just when it is necessary. The data-driven RAJIT leverages the strengths of both optimization methods, leading to favorable outcomes in both safety and economics. Note that the data-driven RAJIT operation scheme is a rolling optimization which is general enough to be employed with deterministic optimization and any data-driven risk-averse operation methods (i.e., the moment-based DRO, the Wasserstein-based DRO, etc.). Without loss of generality, the Online-Learning Distributionally Robust Optimization (OL-DRO) with a learning-informed ambiguity set is implemented. This ambiguity set accurately captures fine-grained characteristics of the uncertainty data based on a Dirichlet process mixture model. Thirdly, the off-grid AHMGs optimization problem is an intractable MINLFP. To address this problem, we develop an equivalent-reformulation-based solution methodology in light of the Glover's linearization and constraint-tightening techniques. The main contributions are summarized as follows.

- A data-driven RAJIT operation scheme is proposed with a just-in-time strategy to maximize cost-effectiveness while hedging against uncertainties. Unlike conventional one-size-fits-all schemes using one optimization method throughout, the proposed scheme strategically switches between two optimization models. Case studies show that this scheme achieves favorable outcomes in both operation safety and economics.
- To enhance the electricity-hydrogen-ammonia coupling, the DGTs and HB processes with tunable feed-in ratios are first modeled and incorporated into the off-grid AHMG. On this basis, a novel state-behavior mapping strategy is developed, considering the practical coordination between multi-mode hydrogen storage and multi-mode ammonia synthesis.
- Based on a proposed constraint-tightening technique and linearization techniques, we develop a tractable reformulation of the proposed MINLFP model to overcome its computational intractability.

The rest of this article is organized as follows. Section II introduces the model formulation. Section III describes the proposed data-driven RAJIT operation scheme. The solution methodology is presented in Section IV. Section V provides case studies. The conclusion is given in Section VI.

II. System Description and Model Formulation

In this section, we begin by providing an overview of the structure of an off-grid AHMG. Then, the mathematical models of each part within the off-grid AHMG are presented in Sections II.A-II.C. Lastly, based on the model of $H_2$ storage and an HB process, the state-behavior mapping strategy is formulated in Section II.D.

As illustrated in Fig. 1, the off-grid AHMG represents a remote industrial park that includes Hydrogen-Ammonia Plants (HAPs), renewable energy sources such as wind turbines and photovoltaic panels, and battery systems. These components are interconnected through a Power Distribution Network (PDN). The HAP incorporates an electrolyzer, a Pressure Swing Adsorption (PSA), an HB process, gas storage systems, and a DGT. Different from the traditional gas turbines which use natural gas as the fuel, the DGT utilizes a mixture of $NH_3$ and $H_2$ as the dual fuel, achieving complete carbon neutrality during operation. The dual fuel is burned in the combustion chamber to drive a gas turbine for electricity generation [38]. The design of DGT is similar to natural gas turbines, resulting in comparable system efficiencies and capital costs. The energy efficiency of DGT further depends on the mixture gas composition, which we will discuss in its model. However, the system setting considered is a zero-carbon off-grid microgrid. To achieve low-carbon generation, traditional natural gas turbines need to be combined with carbon-capture and storage systems, which are of higher investment cost. The HB process is an electro-powered chemical reaction process that mixes $H_2$ and $N_2$ to produce $NH_3$. The whole HB process consists of a mixture and compression module, a reactor, and a separator. Battery is implemented to balance the short-term uncertainties stemming from the forecast errors of renewable energy generation and load [39]. In this section, subscripts of the HAP, battery, and renewable energy are omitted for notational brevity except within the constraints of the PDN.

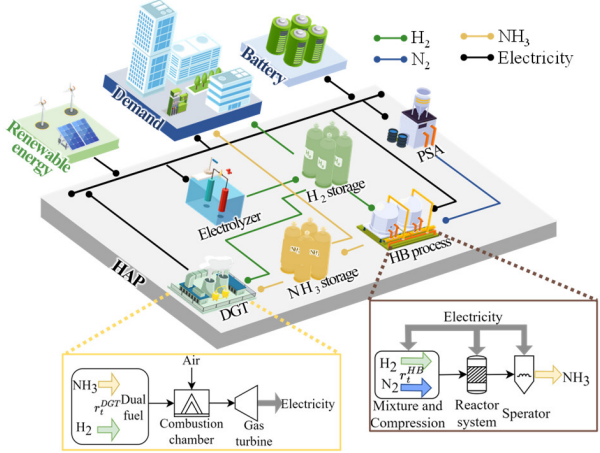

Fig. 1. Structure of an off-grid AHMG.

*E. HAP*

1) Multi-Mode HB Process Modeling

Based on the chemical reaction equilibrium, the material balance is presented in (1). Given that the HB process is energy-intensive, ramp constraints must be enforced, as defined by (2). Eq. (3) restricts that the power of the HB process changes no more than once every $\tau$ hours. According to actual operation conditions, the HB process operation is divided into three modes. Each of the operation modes is assigned with a binary identifier $\varepsilon_{it}^{HB}$, which is defined in (4). The total power load is determined by (5). The feed-in molar ratio is calculated in (6).

$$m_t^{HB} = m_t^{H_2HBin} M^{NH_3} q_t^{NH_3in} \left( M^{H_2} q_t^{H_2in} \right)^{-1}, \forall t \quad (1)$$

$$v_t \delta^- \overline{P}^{HB} \leq P_t^{HB} - P_{t-1}^{HB} \leq v_t \delta^+ \overline{P}^{HB}, \forall t \quad (2)$$

$$\sum_{t'=t-\tau}^{t} v_{t'} \leq 1, \; v_t \in \mathbb{B}, \forall t \quad (3)$$

$$\sum_{i=1}^{3} \varepsilon_{it}^{HB} = 1, \; \varepsilon_{it}^{HB} \in \mathbb{B}, \forall t \quad (4)$$

$$P_t^{HB} = \sum_{i=1}^{3} \varepsilon_{it}^{HB} P_t^{HB-i}, \forall t \quad (5)$$

$$r_t^{HB} = q_t^{H_2in} / q_t^{N_2in}, \forall t \quad (6)$$

In Mode 1, $\varepsilon_{1t}^{HB}$ is assigned to be one. The HB reactor operates with a feed-in $H_2$-to-$N_2$ molar ratio of 3. The power load of HB in Mode 1 is given in (7). The power load of the HB process comes primarily from the reactor and compressor. The power load of the separator is disregarded because it is a minor portion of the overall power consumption [30]. Constraint (8) presents the power load range of Mode 1.

$$\begin{aligned} P_t^{HB-1} &= \eta^{HB} m_t^{H_2HBin} + \eta^{Com} \left( m_t^{N_2in} + m_t^{H_2HBin} \right) \\ &= \left( M^{N_2} \eta^{Com} \left( 3M^{H_2} \right)^{-1} + \eta^{HB} + \eta^{Com} \right) m_t^{H_2HBin}, \forall t \end{aligned} \quad (7)$$

$$\underline{\eta}^{HB} \overline{P}^{HB} \leq P_t^{HB-1} \leq \overline{\eta}^{HB} \overline{P}^{HB}, \forall t \quad (8)$$

In Mode 2, $\varepsilon_{2t}^{HB}$ is assigned to be one. The power load of HB in Mode 2 is presented in (9). The feed-in $H_2$-to-$N_2$ molar ratio range is expanded to allow values between a constant lower bound $\underline{r}^{HB}$ and an upper bound of 3, which is the stoichiometric balanced ratio [40], as presented in (10). With a consistent power load for the HB process, a decrease in the $H_2$-to-$N_2$ molar ratio results in a reduction in $H_2$ intake and $NH_3$ production. Eq. (11) defines the power range in Mode 2.

$$P_t^{HB-2} = \left( M^{N_2} \eta^{Com} \left( M^{H_2} r_t^{HB} \right)^{-1} + \eta^{HB} + \eta^{Com} \right) m_t^{H_2HBin}, \forall t \quad (9)$$

$$\underline{r}^{HB} \leq r_t^{HB} < 3, \forall t \quad (10)$$

$$\underline{\eta}^{HB} \overline{P}^{HB} \leq P_t^{HB-2} \leq \overline{\eta}^{HB} \overline{P}^{HB}, \forall t \quad (11)$$

In Mode 3, $\varepsilon_{3t}^{HB}$ is assigned to be one. The HB process is off in Mode 3, presented by

$$P_t^{HB-3} = 0, \forall t \quad (12)$$

2) DGT Modeling

The $H_2$ blending ratio of the dual fuel is a key variable of the DGT, and it is calculated by the mass of $H_2$ and $NH_3$, as shown in (13). It is well recognized that the combustion efficiency depends on the component of fuels. Accordingly, the effect of the $H_2$ molar ratio in the fuel mixture on the combustion performance in the DGT is represented by a fifth-order polynomial in (14), which is adopted from simulations in [35], and $a_0$-$a_5$ are parameters. Eq. (15) restricts the $H_2$ blending ratio under a upper bound (60%) to avoid combustion instability [35]. Eq. (16) indicates that the unit lower heating value is proportional to the $H_2$ blending ratio [41], where $b_0$-$b_1$ are parameters. The DGT power generation is a product of gas turbine efficiency, combustion efficiency, lower heating value, and the total mass of the fuel, as shown in (17). $H_2$ enjoys a much higher heating value than that of $NH_3$, and therefore the higher the $H_2$ blending ratio is used, the higher the efficiency of DGT is. The range of DGT power is provided in (18).

$$\begin{aligned} r_t^{DGT} &= \frac{q_t^{H_2GTin}}{q_t^{H_2GTin} + q_t^{NH_3GTin}} \\ &= \frac{M^{NH_3} m_t^{H_2GTin}}{M^{NH_3} m_t^{H_2GTin} + M^{H_2} m_t^{NH_3GTin}}, \forall t \end{aligned} \quad (13)$$

$$\begin{aligned} \eta_t^{CE} &= a_5 \left( r_t^{DGT} \right)^5 + a_4 \left( r_t^{DGT} \right)^4 + a_3 \left( r_t^{DGT} \right)^3 \\ &\quad - a_2 \left( r_t^{DGT} \right)^2 + a_1 r_t^{DGT} + a_0, \forall t \end{aligned} \quad (14)$$

$$0 \leq r_t^{DGT} \leq \overline{r}^{DGT}, \forall t \quad (15)$$

$$LHV_t = b_1 r_t^{DGT} + b_0, \forall t \quad (16)$$

$$P_t^{DGT} = \eta^{DGT}\eta_t^{CE} LHV_t \left(m_t^{H_2GTin} + m_t^{NH_3GTin}\right), \forall t \quad (17)$$

$$0 \leq P_t^{DGT} \leq \overline{P}^{DGT}, \forall t \quad (18)$$

*3) H$_2$ Storage Modeling*

The H$_2$ storage dynamics is defined in (19). Drawing upon the concept of hysteresis band [42], the feasible range of the H$_2$ storage State Of Charge (SOC) can be divided into three regions, and each region is represented by a binary indicator $\varepsilon_{it}^{HS}$. A H$_2$ storage can only works in one region at any given time, as defined in (20). The H$_2$ storage level is further described in Eqs. (21)-(22).

$$m_{t+1}^{HS} = m_t^{HS} + \eta^{elz} P_t^{elz} / C_{elz} - m_t^{H_2HBin} - m_t^{H_2D} - m_t^{H_2GTin}, \forall t \quad (19)$$

$$\sum_{i=1}^{3}\varepsilon_{it}^{HS} = 1, \varepsilon_{it}^{HS} \in \mathbb{B}, \forall i = 1,2,3, \forall t \quad (20)$$

$$m_t^{HS} = \sum_{i=1}^{3} m_{it}^{HS}, \forall t \quad (21)$$

$$\varepsilon_{it}^{HS}\eta_i^{HS}\overline{m}^{HS} < m_{it}^{HS} \leq \varepsilon_{it}^{HS}\eta_{i+1}^{HS}\overline{m}^{HS}, \forall t \quad (22)$$

*4) Other Constraints of the HAP*

Eqs. (23)-(24) indicate the dynamics and limitations of NH$_3$ storage, respectively. The load of PSA and HAP are calculated by Eqs. (25)-(26), respectively. The power range of PSA and electrolyzer are shown in Eqs. (27)-(28), respectively.

$$m_{t+1}^{NH_3} = m_t^{NH_3} + m_t^{HB} - m_t^{NH_3GTin} - m_t^{NH_3D}, \forall t \quad (23)$$

$$0 \leq m_t^{NH_3} \leq \overline{m}_t^{NH_3}, \forall t \quad (24)$$

$$P_t^{PSA} = \eta^{PSA} m_t^{N_2in}, \forall t \quad (25)$$

$$P_t^{HAP} = P_t^{elz} + P_t^{HB} + P_t^{PSA} - P_t^{DGT}, \forall t \quad (26)$$

$$0 \leq P_t^{PSA} \leq \overline{P}^{PSA}, \forall t \quad (27)$$

$$0 \leq P_t^{elz} \leq \overline{P}^{elz}, \forall t \quad (28)$$

*F. Battery*

Eq. (29) presents the battery dynamics. The battery charging and discharging powers are constrained by (30)-(31), respectively. The battery cannot be charged and discharged at the same time, as enforced by (32). The storage level constraints are presented in (33).

$$e_{t+1}^{BT} = e_t^{BT} + \eta_b^{ch} P_t^{ch} - P_t^{dch}/\eta_b^{dch}, \forall t \quad (29)$$

$$0 \leq P_t^{ch} \leq \varepsilon_t^{ch}\overline{P}^{BT}, \forall t \quad (30)$$

$$0 \leq P_t^{dch} \leq \varepsilon_t^{dch}\overline{P}^{BT}, \forall t \quad (31)$$

$$\varepsilon_t^{ch} + \varepsilon_t^{dch} \leq 1, \varepsilon_t^{ch}, \varepsilon_t^{dch} \in \mathbb{B}, \forall t \quad (32)$$

$$\underline{\eta}^{BT}\overline{E}^{BT} \leq e_t^{BT} \leq \overline{\eta}^{BT}\overline{E}^{BT}, \forall t \quad (33)$$

*G. Power Distribution Network*

The PDN in the off-grid AHMG must adhere to power balance constraints known as the linearized DistFlow model, as presented in Eqs. (34)-(38).

$$\sum_{w \in S^{WT}} s_{jw}^{WT}\{\}_{tw}^{WT} + \sum_{w \in S^{PV}} s_{jw}^{PV}\{\}_{tw}^{PV} + \sum_{b \in S^{BT}} s_{jb}^{BT}\left(\{\}_{tb}^{dch} - \{\}_{tb}^{ch}\right) +$$
$$\sum_{(i,j) \in S^L}\{\}_{t(i,j)}^{line} = \sum_{(j,k) \in S^L}\{\}_{t(j,k)}^{line} + \sum_{p \in S^{HAP}}s_{jp}^{HAP}\{\}_{tp}^{HAP} + \{\}_{tj}^{L} - \{\}_{tj}^{C}, \forall t \quad (34)$$

$$-\{\}_{t(j,k)}^{max} \leq \{\}_{t(j,k)}^{line} \leq \{\}_{t(j,k)}^{max}, \{\} = P,Q, \forall t, (j,k) \quad (35)$$

$$U_{tj} - U_{tk} = \left(R_{(j,k)}P_{t(j,k)}^{line} + X_{(j,k)}Q_{t(j,k)}^{line}\right)/U_0, \forall t,j,k \quad (36)$$

$$U_j^{\min} \leq U_{tj} \leq U_j^{\max}, \forall t, j \quad (37)$$

$$Q_{tj}^{L} = P_{tj}^{L}\tan\left(\cos^{-1}\theta_j\right), \forall t, j \quad (38)$$

*H. State-Behavior Mapping Strategy*

The H$_2$ for ammonia synthesis is obtained from H$_2$ storage, so it is reasonable to determine the operation mode of the HB process according to the H$_2$ storage level. Given this observation, a novel state-behavior mapping strategy is proposed and illustrated in Fig. 2, regarding the SOC of H$_2$ storage and the behavior (operation mode) of the HB process. As shown in region ① of Fig. 2, when the H$_2$ storage SOC is situated in $\left[\eta_1^{HS}, \eta_2^{HS}\right]$ (i.e. $\varepsilon_{3t}^{HS} = 1$), which is a hysteresis band near the lower bound, the over-low H$_2$ storage level may incur unanticipated shutdowns. These shutdowns can damage the reactor and catalysts and lead to a reduction in the lifespan. Therefore, logical constraint (39) enforces the HB process to shut down in a scheduled manner for safety concerns. When the H$_2$ storage SOC is in $\left[\eta_2^{HS}, \eta_3^{HS}\right]$ (i.e. $\varepsilon_{2t}^{HS} = 1$), logical constraint (40) indicates that the HB process can be in all three operation modes, as illustrated in region ②. When the SOC of H$_2$ storage is in $\left[\eta_3^{HS}, \eta_4^{HS}\right]$ (i.e. $\varepsilon_{1t}^{HS} = 1$), signifying the abundance of H$_2$, the HB process can either be off (Mode 3) or work at a maximum feed-in ratio (Mode 1) to maximize productivity and energy efficiency [40], as given in Eq. (41) and region ③ of Fig. 2. Eq. (42) restricts that the start-up from Mode 3 to Mode 2 is not permitted. In other words, the HB process can only be started with the state transition from Mode 3 to Mode 1, ensuring that it is well-prepared to operate at a stoichiometric balanced ratio, as presented in arrow ④. The shutdown from Mode 1 to Mode 3 is not allowed, meaning that the HB process can only be shut down from Mode 2 to Mode 3. This ensures that the shutdown occurs when the HB process cannot maintain at the stoichiometric balanced ratio, as presented in Eq. (43) and arrow ⑤.

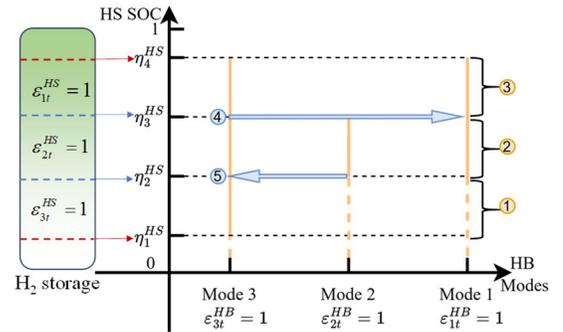

Fig. 2. The state-behavior mapping strategy. The blue arrows indicate the possible transitions between modes.

$$\varepsilon_{3t}^{HS} = 1 \Rightarrow \varepsilon_{3t}^{HB} = 1, \forall t \quad (39)$$

$$\varepsilon_{2t}^{HS} = 1 \Rightarrow \varepsilon_{1t}^{HB} + \varepsilon_{2t}^{HB} + \varepsilon_{3t}^{HB} = 1, \forall t \quad (40)$$

$$\varepsilon_{1t}^{HS} = 1 \Rightarrow \varepsilon_{1t}^{HB} + \varepsilon_{3t}^{HB} = 1, \forall t \quad (41)$$

$$\varepsilon_{2(t+1)}^{HB} = 1 \Rightarrow \varepsilon_{3t}^{HB} = 0, \forall t \quad (42)$$

$$\varepsilon_{3(t+1)}^{HB} = 1 \Rightarrow \varepsilon_{1t}^{HB} = 0, \forall t \quad (43)$$

## III. Mathematical Formulation of RAJIT Operation Scheme

In this section, the proposed RAJIT operation scheme is presented based on the off-grid AHMG model formulated in Section II. The schematic diagram is illustrated in Fig. 3. Uncertainty data are continuously collected into an online dataset and then used to update a learning-informed ambiguity set. An indicator named *flag* is employed to track the actual risk profile of the off-grid AHMG. Then, the data-driven RAJIT switches the optimization strategy between the deterministic optimization and the OL-DRO according to the value of *flag*, ensuring a just-in-time response to the actual risk profile.

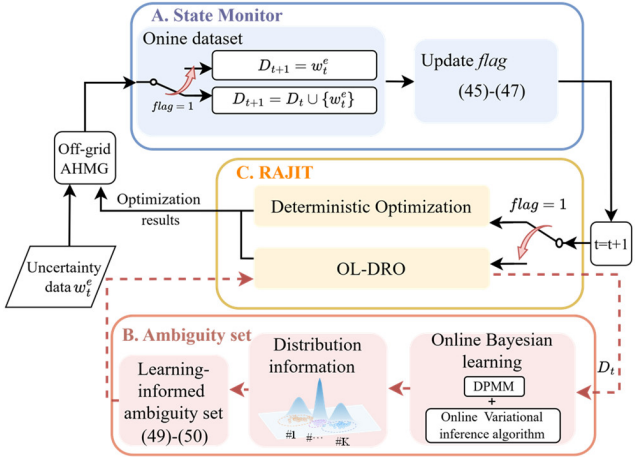

Fig. 3. Schematic diagram of the proposed RAJIT operation scheme.

### A. State Monitor

The state monitor updates the online dataset and captures the actual risk profile. The indicator *flag* serves a dual purpose. It determines the update of the online dataset, and it is instrumental in deciding the switch of optimization methods.

1) Update Online Dataset

The online dataset $\mathcal{D}_t$ is retained for the OL-DRO, and is updated after a new uncertainty data point $w_t^e$ is collected. A *flag* value of 1 signifies that the underlying information in the current online dataset has been captured by the online Bayesian learning, and is subsequently incorporated into the learning-informed ambiguity set. Therefore, this dataset is discarded for memory saving, and $w_t^e$ is preserved individually as the new online dataset. Otherwise, the new data is appended to the existing dataset. This update rule is summarized and formulated in (44).

$$D_0 = \varnothing,$$
$$D_{t+1} = flag \cdot \{w_t^e\} \oplus (1-flag) \cdot D_t \cup \{w_t^e\}, \ \forall t \geq 1 \quad (44)$$

where $\oplus$ denotes the Minkowski sum.

2) Update Flag

The indicator *flag* is a binary parameter which indicates whether the off-grid AHMG is predicted to be unsafe in the following $l_{ET}$ time periods, as depicted in Fig. 4. $l_{ET}$ is a manually specified parameter that controls the sensitivity to detect unsafety. The *flag* is updated at each time period, and the logic is given by (45).

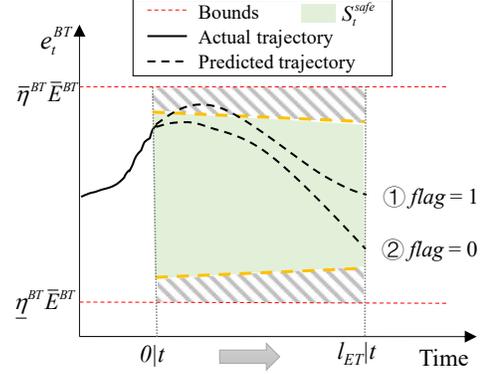

Fig. 4. Illustration of the update logic of the *flag*.

$$flag = \begin{cases} 0, & \text{if } \mathbf{e}_t^{BT} \in S_t^{safe} \\ 1, & \text{otherwise} \end{cases} \quad (45)$$

where $\mathbf{e}_t^{BT}$ denotes the matrix of $l_{ET}$-step ahead battery storage level prediction, as defined in (46).

$$\mathbf{e}_t^{BT} = \begin{bmatrix} e_{1,1|t}^{BT} & \cdots & e_{1,l_{ET}|t}^{BT} \\ \vdots & \ddots & \vdots \\ e_{|S^{BT}|,1|t}^{BT} & \cdots & e_{|S^{BT}|,l_{ET}|t}^{BT} \end{bmatrix}, \forall t \quad (46)$$

The safe region $S_t^{safe}$ is a set of battery storage levels, and this region is slightly smaller than the rated operating range, as presented by (47).

$$S_t^{safe} = \left\{ \mathbf{e}_t^{BT} \left| \begin{bmatrix} \mathbf{I}^{|S^{BT}|} \\ -\mathbf{I}^{|S^{BT}|} \end{bmatrix} \mathbf{e}_t^{BT} \leq \begin{bmatrix} \overline{\eta}^{BT} \overline{E}^{BT} \mathbf{1}^{|S^{BT}| \times l_{ET}} \\ -\underline{\eta}^{BT} \overline{E}^{BT} \mathbf{1}^{|S^{BT}| \times l_{ET}} \end{bmatrix} \right. \right.$$
$$\left. + \frac{\eta_b^{ch} \overline{\omega}}{|S^{BT}|} \begin{bmatrix} -\mathbf{A} \\ \mathbf{A} \end{bmatrix} \right\} \quad (47)$$

where $\mathbf{1}$ and $\mathbf{I}$ are the full one matrix and the identity matrix, respectively. $\mathbf{A} \in \mathbb{R}^{|S^{BT}| \times l_{ET}}$ is a matrix with each element $a_{ij}=j$. $\oplus$ and $\ominus$ are the Minkowski set addition and the Pontryagin set difference, respectively. $\mathbb{W}^e = \left\{ w \left| \begin{bmatrix} 1 & -1 \end{bmatrix}^T w \leq \begin{bmatrix} \overline{\omega} & \overline{\omega} \end{bmatrix} \right. \right\}$ is the predefined support set of uncertainty data.

### B. Learning-Informed Ambiguity Set

The online streaming uncertainty data continue being fed into the system at each time slot, as shown in Fig. 3. As time goes by, the structure of dataset can be time-varying. To automatically capture the underlying structure of the uncertainty data distribution from the online dataset $D_t$, we employ a nonparametric Bayesian model known as the Dirichlet process mixture model along with an online variational inference algorithm. Then, the online variational inference results are leveraged to construct the learning-informed ambiguity set, which encapsulates the multimodality and local moment information of uncertainties.

The Dirichlet process mixture model is described as a widely used stick-breaking representation of the Dirichlet process.

$$G = \sum_{k=1}^{\infty} \pi_k \delta_{\theta_k}, \quad \pi_k = V_k \prod_{i=1}^{k-1}(1-V_i)$$
$$V_i \sim Beta(1,\alpha), \quad w_i^e | \zeta_i \sim F(\theta_{\zeta_i})$$
(48)

where $G$ is a random distribution generated through the stick-breaking process, which ensures that $G$ is discrete with probability one. This discreteness naturally induces clustering in the data, as each data point is assigned to a specific mixture component. $\alpha$ is a scaling parameter. $\zeta_i$ denotes a latent variable of the $i$-th uncertainty data, indicating that this data point belongs to the $\zeta_i$-th mixture component. $\theta_{\zeta_i}$ denotes a set of data distribution parameters of the $\zeta_i$-th component. $V_i$ is the proportion of a broken stick from the remaining stick, following a Beta distribution. $F(\cdot)$ is the Gaussian distribution.

Then, an online variational inference algorithm is employed to approximate the conditional distribution of latent variables with observed data [43], [44]. This algorithm iterates between a model-building phase and a memory compression phase [45]. The online variational inference results provide the multimodal statistical information of the uncertainty data, including the weight $\gamma_i$, the mean $\mu_i$, and the covariance $\Sigma_i$ of the $i$-th component. Accordingly, we formulate the learning-informed ambiguity set $\mathcal{P}$ as a weighted Minkowski sum of $K$ basic ambiguity sets $\mathcal{P}_i$, as presented in (49)-(50).

$$\mathcal{P} = \sum_{i=1}^{K} \gamma_i \mathcal{P}_i (\mathbb{W}^e, \mu_i, \Sigma_i)$$ (49)

$$\mathcal{P}_i(\mathbb{W}^e, \mu_i, \Sigma_i) = \left\{ \theta \in \mathcal{M}_+ \middle| \begin{array}{l} \int_{\mathbb{W}^e} \theta(d\xi) = 1, \\ \int_{\mathbb{W}^e} \xi \theta(d\xi) = \mu_i, \\ \int_{\mathbb{W}^e} \xi \xi^T \theta(d\xi) \leq \Sigma_i + \mu_i \mu_i^T \end{array} \right\}$$ (50)

The learning-informed ambiguity set learns the fine-grained local information of the uncertainty data from the online dataset automatically, and therefore it outperforms existing DRO methods.

*C. Data-Driven RAJIT Optimization Problem*

Based on the learning-informed ambiguity set, we present the formulation of the data-driven RAJIT. It includes two problems, namely deterministic optimization and OL-DRO. If *flag*=1, the operation safety is the primary concern. Therefore, it is imperative for the off-grid AHMG to switch the optimization strategy to the OL-DRO to be risk-averse. Otherwise, the off-grid AHMG switches back to deterministic optimization for cost-saving.

In the RAJIT operation scheme, the deterministic optimization problem is summarized as follows.

$$\min J_t = \sum_{l=0}^{N_p-1} \left\{ \underbrace{\sum_{i \in \Omega_b} c^C P_{i,l|t}^C}_{C_{l|t}^C} + \underbrace{\left(c^{BT} e_{l|t}^{BT} + c^{HS} m_{l|t}^{HS}\right)}_{C_{l|t}^{de}} \right.$$
$$\left. + \underbrace{c^{ess} \sum_{i=1}^{|S^{BT}|}\left(P_t^{ch}+P_t^{dch}\right) + c^{DGT} P_{l|t,i}^{DGT} + c^{elz} P_{l|t}^{elz} + c_{AD}^{HB} v_{l|t}}_{} \right.$$
$$\left. + \underbrace{c_M^{HB} P_{l+1|t}^{HB} + c_{SU}^{HB}\left(\varepsilon_{3,l+1|t}^{HB}, \varepsilon_{1,l+1|t}^{HB}\right) + c_{SD}^{HB}\left(\varepsilon_{2,l|t}^{HB}, \varepsilon_{3,l+1|t}^{HB}\right)}_{C_{l|t}^{op}} \right\}$$ (51)

s.t. HAP operation constraints (1)-(28)

Battery operation constraints (29)-(33)
Power distribution network constrains (34)-(38)
State-behavior mapping strategy (39)-(43)

where $l|t$ denotes the $l$-step ahead from the sampling time $t$.

The OL-DRO problem is the same as the deterministic optimization problem, except that the nominal constraint (33) on battery storage level is replaced by the risk-averse Distributionally Robust Conditional-Value-at-Risk (DR-CVaR) constraints defined in (52)-(53) [46]. In deterministic optimization with nominal constraints on battery storage levels, the actual source-load uncertainty may result in the battery SOC frequently exceeding the safety bounds. Instead, the DR-CVaR constraints regulate the constraint violation of battery storage level in both frequency and magnitude.

$$\sup_{\mathbb{P} \in \mathcal{P}} \mathbb{P}\text{-CVaR}_\varepsilon \left( \begin{array}{l} e_t^{BT} + \eta_b^{ch} P_t^{ch} - P_t^{dch}/\eta_b^{dch} \\ + w_t^e - \overline{\eta}^{HS} \overline{E}^{BT} \end{array} \right) \leq 0, \forall t$$ (52)

$$\sup_{\mathbb{P} \in \mathcal{P}} \mathbb{P}\text{-CVaR}_\varepsilon \left( \begin{array}{l} \underline{\eta}^{HS} \overline{E}^{BT} - e_t^{ch} - \eta_b^{ch} P_t^{BT} + \\ P_t^{dch}/\eta_b^{dch} - w_t^e \end{array} \right) \leq 0, \forall t$$ (53)

where $\varepsilon$ denotes the given tolerance level for the DR-CVaR constraint. $\mathcal{P}$ is the learning-informed ambiguity set. Following [44], the DR-CVaR constraint is further defined as follows:

**Definition (DR-CVaR):** For a given measurable loss $L$: $\mathbb{R}^n \to \mathbb{R}$, a random vector $\tilde{\xi}$ on $\mathbb{R}^n$ with probability $\mathbb{P}$, and tolerance level $\varepsilon$, the CVaR of random loss function $L$ at level $\varepsilon$ with respect to any probability $\mathbb{P}$ in ambiguity set $\mathcal{P}$ is defined below.

$$\sup_{\mathbb{P} \in \mathcal{P}} \mathbb{P} - CVaR_\varepsilon\left(L(\tilde{\xi})\right) = \sup_{\mathbb{P} \in \mathcal{P}} \inf_{\beta \in \mathbb{R}} \left\{ \beta + \frac{1}{\varepsilon} \mathbb{E}_{\mathbb{P}}\left[\left(L(\tilde{\xi}) - \beta\right)^+\right] \right\}$$ (54)

where $\mathbb{E}_{\mathbb{P}}$ denotes the expectation with respect to probability $\mathbb{P}$. $\beta$ is an auxiliary variable. The DR-CVaR can be interpreted as the conditional expectation of loss $L$ above the $1-\varepsilon$ quantile of the probability distribution $\mathbb{P}$ in ambiguity set $\mathcal{P}$.

The OL-DRO problem is given as follows.
  min Objective function in (51)
  s.t. HAP operation constraints (1)-(28)
    Battery operation constraints (29)-(32)
    Power distribution network constrains (34)-(38)
    State-behavior mapping strategy (39)-(43)
    DR-CVaR constrains (52)-(53)
    Ambiguity set (49)-(50)

By leveraging the proposed RAJIT scheme, the off-grid AHMG intelligently and automatically switches the real-time optimization strategy just in time between risk-agnostic deterministic optimization and risk-averse OL-DRO according to its actual risk profile. When SOC predictions enter unsafe regions, OL-DRO is employed to minimize constraint violation rates with the least cost. When SOC predictions remain within safe regions, deterministic optimization is employed to maximize economic efficiency, capitalizing on its superior cost-effectiveness. Therefore, the strengths of both optimization methods are fully harnessed, yielding favorable operation performance in terms of both safety and economics. Furthermore, the adjustable confidence level parameter during OL-DRO operation provides a flexible mechanism for precisely calibrating the safety-economic performance trade-off, enabling fine-tuned operational control under varying system operation conditions.

**Remark:** It is important to clarify that our study focuses on look-ahead rolling operation with an hourly time resolution. Based on existing literature on multiscale scheduling/operation optimization [37], we can extend our proposed method for multiple time-scale coordination. Specifically, on annual/monthly scales, we can determine future hydrogen and ammonia inventory management schemes. Our proposed method can then be applied for real-time adjustments. Finer-grained real-time control can be implemented based on the scheduling solutions provided by our method.

## IV. PROPOSED SOLUTION METHODOLOGY

There are two challenges when solving the formulated data-driven RAJIT problem. First, the problem is an intractable MINLFP because of the nonlinear constraints in the off-grid AHMG model. Therefore, reformulation techniques are introduced in Sections IV.A-C. Second, the learning-informed ambiguity in the DR-CVaR constraints poses a computational challenge. To address this problem, a constraint-tightening technique is developed in Section IV.D.

### A. Linearization of Multi-Mode HB Process

Eq. (5) is nonlinear due to the product terms of binary variables and continuous variables. These terms are linearized using the Glover's linearization method, presented by (55).

$$P_t^{HB-i} - M\left(1 - \varepsilon_{it}^{HB}\right) \leq P_t^{HB} \leq P_t^{HB-i} + M\left(1 - \varepsilon_{it}^{HB}\right), \quad (55)$$
$$\forall t, \forall i = 1, 2, 3$$

### B. Reformulation of State-Behavior Mapping Strategy

The state-behavior mapping strategy in Eqs. (41)-(43) cannot be directly implemented in the model, so they are equivalently reformulated as follows.

$$\varepsilon_{1t}^{HS} \leq \varepsilon_{1t}^{HB} + \varepsilon_{3t}^{HB}, \; \varepsilon_{2t}^{HS} \leq \varepsilon_{1t}^{HB} + \varepsilon_{2t}^{HB} + \varepsilon_{3t}^{HB}, \; \varepsilon_{3t}^{HS} \leq \varepsilon_{3t}^{HB}, \forall t \quad (56)$$

$$\varepsilon_{2(t+1)}^{HB} \leq 1 - \varepsilon_{3t}^{HB}, \forall t \quad (57)$$

$$\varepsilon_{3(t+1)}^{HB} \leq 1 - \varepsilon_{1t}^{HB}, \forall t \quad (58)$$

### C. Reformulation of Fractional Terms and Multilinear Terms

To address the fractional term in Eq. (9), we introduce an auxiliary variable $z_t$, as defined in Eq. (59). Consequently, Eq. (9) is reformulated as Eq. (60).

$$z_t = 1/r_t^{HB}, \forall t \quad (59)$$

$$P_t^{HB-2} = \left[\left(z_t M^{N_2}/M^{H_2} + 1\right)\eta^{Com} + \eta^{HB}\right]m_t^{H_2HBin}, \forall t \quad (60)$$

In addition, the fractional term in (13) causes intractability to the problem. An equivalent bilinear reformulation is obtained with an auxiliary variable $m_t$ below.

$$r_t^{DGT} = M^{NH_3} m_t^{H_2GTin} m_t, \forall t \quad (61)$$

$$m_t\left(M^{NH_3} m_t^{H_2GTin} + M^{H_2} m_t^{NH_3GTin}\right) = 1, \forall t \quad (62)$$

Moreover, the multilinear terms in (14) can be equivalently reformulated as multiple bilinear terms. In specific, we use $r$ to represent term $r_t^{DGT}$. Then, we have $r^5 = r(r^4)$, $r^4 = (r^2)^2$, and $r^3 = r(r^2)$. Define lifted auxiliary variable $s$ and $p$ as follows

$$s = r^2, \; p = s^2 \quad (63)$$

such that we have

$$r^3 = rs, \; r^4 = s^2, \; r^5 = rp. \quad (64)$$

By the above formulations, we derive the equivalent reformulation of (14) below.

$$\eta_t^{CE} = a_5 rp + a_4 s^2 + a_3 rs - a_2 s + a_1 r_t^{DGT} + a_0, \forall t \quad (65)$$

The corresponding quadratically constrained program including bilinear terms can be directly solved by off-the-shelf solvers such as GUROBI.

### D. Constraint-Tightening Technique

The purpose of the constraint tightening is to equivalently replace the original intractable DR-CVaR constraints with hard-tightened constraints.

Since the DR-CVaR constraint needs to be satisfied for $\mathbf{e}_t^{BT}$ and any $t$ in the horizon, we leverage the tube-based method [43]. (52)-(53) can be cast as follows.

$$\overline{\mathbf{e}}_t^{BT} \in \left\{ \mathbf{e}_t^{BT} \middle| \begin{array}{l} \left[\begin{array}{c} \mathbf{I}^{|S^{BT}|} \\ -\mathbf{I}^{|S^{BT}|} \end{array}\right] \mathbf{e}_t^{BT} \leq \left[\begin{array}{c} \overline{\eta}^{BT} \overline{E}^{BT} \mathbf{1}^{|S^{BT}| \times l_{ET}} \\ -\underline{\eta}^{BT} \overline{E}^{BT} \mathbf{1}^{|S^{BT}| \times l_{ET}} \end{array}\right] \\ + \dfrac{\eta_b^{ch} \overline{\omega}}{|S^{BT}|} \left[\begin{array}{c} -\left(\mathbf{A} - \mathbf{1}^{|S^{BT}| \times l_{ET}}\right) \\ \left(\mathbf{A} - \mathbf{1}^{|S^{BT}| \times l_{ET}}\right) \end{array}\right] - \boldsymbol{\eta} \mathbf{1}^{1 \times l_{ET}} \end{array} \right\}, \forall t \quad (66)$$

where $\boldsymbol{\eta} \in \mathbb{R}^{2|S^{BT}| \times 1}$ is a vector of constraint tightening terms. Following [43], $\boldsymbol{\eta}$ can be determined online by solving the following problem repeatedly.

$$\min \; \eta_j$$

$$\text{s.t.} \; \inf_{\beta_j \in \mathbb{R}} \left\{ \beta_j + \frac{1}{\varepsilon} \sup_{\mathbb{P} \in \mathcal{P}} \mathbb{E}_{\mathbb{P}} \left\{ \left( \left[\begin{array}{c} \mathbf{I}^{|S^{BT}|} \\ -\mathbf{I}^{|S^{BT}|} \end{array}\right]_j \mathbf{1}^{|S^{BT}| \times 1} \frac{\eta_b^{ch}}{|S^{BT}|} w_t^e \\ -\eta_j - \beta_j \end{array} \right)^+ \right\} \right\} \leq 0 \quad (67)$$

where $[\mathbf{x}]_j$ denotes the $j$-th row of the matrix $\mathbf{x}$. The worst expectation term is further rewritten as follows based on the definition of ambiguity set in (49)-(50).

$$\sup \sum_{i=1}^{K} \gamma_i \int_{\mathbb{W}^e} \left( \left[\begin{array}{c} \mathbf{I}^{|S^{BT}|} \\ -\mathbf{I}^{|S^{BT}|} \end{array}\right]_j \mathbf{1}^{|S^{BT}| \times 1} \frac{\eta_b^{ch}}{|S^{BT}|} w_t^e - \eta_j - \beta_j \right)^+ \theta(d\xi)$$

$$\int_{\mathbb{W}^e} \theta(d\xi) = 1, \quad (68)$$

$$\text{s.t.} \int_{\mathbb{W}^e} \xi \theta(d\xi) = \mu_i, \quad \forall i = 1, \ldots K$$

$$\int_{\mathbb{W}^e} \xi \xi^T \theta(d\xi) \leq \Sigma_i + \mu_i \mu_i^T$$

By taking the dual of optimization problem, (68) is reformulated as follows.

$$\min_{t, \omega_i, \Omega_i} \sum_{i=1}^{K} \gamma_i \left\{ t + \mu_i^T \omega_i + \left(\Sigma_i + \mu_i \mu_i^T\right) \Omega_i \right\} \quad (69)$$

$$\text{s.t.} \quad t + \xi^T \omega_i + \xi^T \Omega_i \xi \geq 0, \quad \forall \xi \in \mathbb{W}^e \tag{70}$$

$$-\left( \begin{bmatrix} \mathbf{I}^{|S^{BT}|} \\ -\mathbf{I}^{|S^{BT}|} \end{bmatrix}_j \mathbf{1}^{|S^{BT}| \times 1} \frac{\eta_b^{ch}}{|S^{BT}|} w_t^e - \eta_j - \beta_j \right) \tag{71}$$

$$+ t + \xi^T \omega_i + \xi^T \Omega_i \xi \geq 0, \quad \forall \xi \in \mathbb{W}^e$$

$$\Omega_i \succeq 0, \quad \forall i = 1,...K \tag{72}$$

The inequality constraints (70)-(71) can be further reformulated into second-order cone constraints in light of duality theory and Schur complements [43], as given in (73)-(74), respectively.

$$\left\| \begin{matrix} \omega_i + E^T \varphi_i \\ \Omega_i - t_i + f^T \varphi_i \end{matrix} \right\|_2 \leq \Omega_i + t_i - f^T \varphi_i \tag{73}$$

$$\left\| \begin{matrix} \omega_i - \frac{\eta_b^{ch}}{|S^{BT}|} \begin{bmatrix} \mathbf{I}^{|S^{BT}|} \\ -\mathbf{I}^{|S^{BT}|} \end{bmatrix}_j \mathbf{1}^{|S^{BT}| \times 1} + E^T \phi_i \\ \Omega_i - t_i - \beta_j - \eta_j + f^T \phi_i \end{matrix} \right\|_2 \leq \begin{matrix} \Omega_i + t_i + \beta_j \\ + \eta_j - f^T \phi_i \end{matrix} \tag{74}$$

where $t_{ij}$, $\omega_{ij}$, and $\Omega_{ij}$ are dual decision variables for the constraints defining the *i-th* basic ambiguity set of the *j-th* DR-CVaR. $\varphi_{ij}, \phi_{ij} \geq 0$ are dual decision variables for the constraints in the support set $\mathbb{W}^e$.

By implementing the above reformulations, the original MINLFP can be recast into a math program with conic constraints which can be directly solved by off-the-shelf solvers.

## IV. CASE STUDIES

In this section, the proposed RAJIT operation scheme is implemented on two off-grid AHMG systems, aiming to demonstrate its effectiveness, superiority, and scalability. We begin by conducting numerical comparative cases on a modified IEEE 33-bus system, highlighting the superior performance of the proposed scheme. To further assess the scalability, we employ a modified IEEE 69-bus system.

TABLE II. KEY PARAMETERS IN THE CASE STUDY

| Parameter | Value | Parameter | Value | Parameter | Value |
|---|---|---|---|---|---|
| $\varepsilon$ | 15% | $c_M^{HB}$ [$/kWh] | 0.076 | $\underline{r}^{HB}$ | 0.418 [40] |
| $c^C$ [$/kWh] | 0.1 | $c_{AD}^{HB}$ [$] | 5 | $\eta^{HB}/\overline{\eta}^{HB}$ | 0.2/1 |
| $c^{BT}$ [$/kWh] | 0.001 | $c_{SU}^{HB}, c_{SD}^{HB}$ [$] | 200 | $\overline{r}^{DGT}$ | 0.6 |
| $c^{HS}$ [$/kg] | 0.0031 | $\delta^+, \delta^-$ | 0.2 | $\eta^{DGT}$ | 0.4 |
| $c^{ess}$ [$/kWh] | 0.0005 | $\overline{P}^{HB}$ [kW] | 2000 | $\overline{P}^{DGT}$ [kW] | 100 |
| $c^{DGT}$ [$/kWh] | 0.026 | $\tau$ [h] | 4 | $\eta^{elz}$ | 0.9 |
| $c^{elz}$ [$/kWh] | 0.01 | $M^{N_2}/M^{H_2}$ $/M^{NH_3}$ [g/mol] | 14/2/10 | $C_{elz}$ [kWh/kg] | 41.97 |
| $\eta_{1,2,3,4}^{HS}$ | 0.1,0.2, 0.8,0.9 [47] | $\eta_b^{ch}/\eta_b^{dch}$ | 0.9/0.9 | $\overline{P}^{BT}$ [kW] | 100 |
| $\overline{E}^{BT}$ [kWh] | 1000 | $\overline{\eta}^{BT}/\underline{\eta}^{BT}$ | 0.9/0.2 | $N_p$ [h] | 24 |

### A. Benchmarking Cases and Comparison Results

In subsection, we employ a modified IEEE 33-bus system, as illustrated in Fig. 5. Therefore, without loss of generality, WTs are considered in this model as distributed renewable generators. It is equipped with three wind turbines located at Buses 2, 11, and 28, two battery systems positioned at Buses 5 and 10, and an HAP located at Bus 8.

Five case studies are set up and listed in TABLE III. Specifically, Case 4 is the proposed scheme. Cases 1, 2, and 4 are designed for comparison to show the effect of the flexible DGT. Case 3 and Case 4 are designed for comparison to clarify the advantages of the state-behavior mapping strategy. The overall cost breakdown under these four cases is illustrated in Fig. 6.

TABLE III. CASE DESCRIPTION

| Case No. | Optimization method | DGT | State-behavior mapping |
|---|---|---|---|
| 1 | RAJIT+OL-DRO | ×($NH_3$-only gas turbine) | √ |
| 2 | RAJIT+OL-DRO | ×(Traditional natural gas turbine) | √ |
| 3 | RAJIT+OL-DRO | √ | × |
| **4** | **RAJIT+OL-DRO** | **√** | **√** |

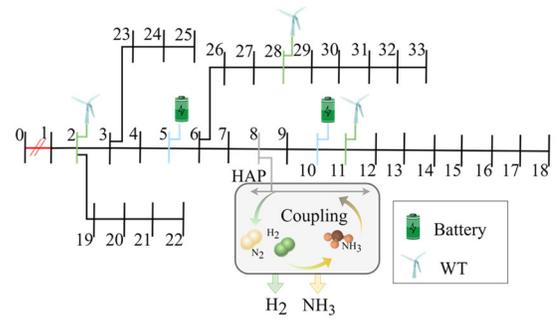

Fig. 5. Topology of the IEEE 33-bus distribution system with off-grid AHMG. The gradient arrows in the HAP indicate multi-energy conversion.

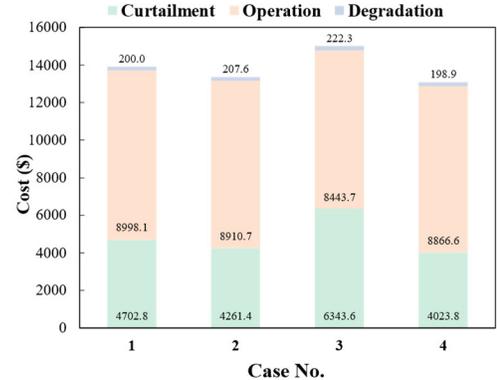

Fig. 6. Overall cost breakdown under different cases.

#### 1) Comparative Analysis on the Effect of DGT

We compare the computational results of Case 1, 2, and 4 to assess the performance of the off-grid AHMG with the DGT. In Cases 1 and 2, the conventional $NH_3$-only gas turbine and natural gas-fueled gas turbine are implemented instead of the DGT, respectively. By contrast, Case 4 employs the DGT with a flexible $H_2$ blending ratio. The detailed operation condition of the DGT in Case 4 is present in Fig. 7.

As illustrated in Fig. 7, the power generation of DGT is concentratedly high at 18:00-19:00 and 22:00-24:00. This is because these periods coincide with both high-power loads and high HB process power. As a backup power generator, the DGT operates strategically during source-load imbalance periods. The DGT operates only when necessary, with a maximum output power of 100kW and high flexibility. Moreover, the $H_2$ blending ratio diminishes to nearly zero during these periods with low hydrogen storage, reflecting an intentional $H_2$ stockpile. During periods characterized by high hydrogen storage level, the $H_2$ blending ratio surges to as high as 0.6 to improve DGT efficiency. In contrast to Case 1, where the $H_2$ blending ratio remains constant at zero, the deployment of a DGT in Case 4 provides extra flexibility, and allows for a more efficient allocation of green $H_2$ and $NH_3$. Consequently, Case 4 achieves an enhanced electricity-hydrogen-ammonia coupling, leading to a 6.2% reduction in overall cost compared to Case 1, as shown in Fig. 6. In Case 2, the reliance on a natural gas-fueled turbine, coupled with a natural gas price of $0.46/m³, leads to significantly higher operational expenses than that of Case 4, and the overall cost is higher than that of Case 4 by 1.0%.

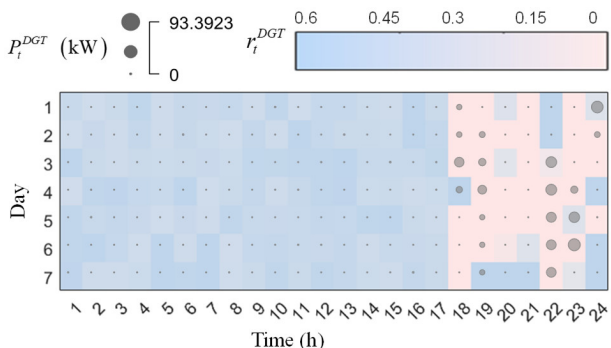

Fig. 7. $H_2$ blending ratio and power generation of the DGT in Case 9.

2) Comparative Analysis on the Effect of Proposed State-Behavior Mapping

We compare the results of Case 3 and Case 4 to showcase the superiority of the state-behavior mapping strategy. In Case 3, we employ the conventional HB process model with a fixed feed-in $H_2$-to-$N_2$ ratio of 3:1 [16], [17]. This leads to merely two operation modes of the HB process, namely on and off. Moreover, the power load of the HB process is assumed to be simply proportional to the amount of feed-in $H_2$.

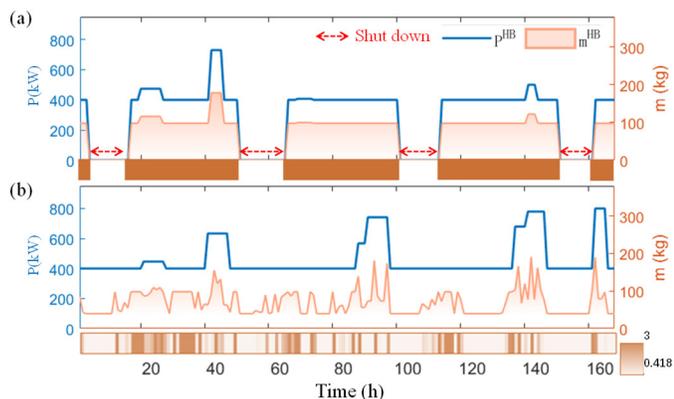

Fig. 8. Results of the HB process of (a). Case 3, (b). Case 4. The area graphs correspond to the right y-axis and the line graphs correspond to the left y-axis. The heat maps under each subplot denote the corresponding feed-in ratio $r^{HB}$. When the HB process is shut down, there is no feed-in ratio $r^{HB}$.

We illustrate the operation conditions of the HB process in Case 3 and Case 4 in Fig. 8. In Case 3, the HB process fails to adjust its operation power due to insufficient green $H_2$ storage, resulting in four undesirable shutdowns at the 5-*th*, 50-*th*, 100-*th*, and 152-*nd* hour, respectively. As a result, the shutdowns of the HB process in Case 3 not only leads to a high operation cost ($8,443.7) associated with shutdowns and startups, but also results in a significant wind energy curtailment cost, amounting to $6,343.6. By contrast, we can see from Fig. 8. (b) that the feed-in $H_2$-to-$N_2$ ratio can be flexibly adjusted ranging from 0.418 to 3 to coordinate with the $H_2$ storage level, and therefore avoiding undesirable shutdowns. During periods characterized by low $H_2$ storage levels and high operation power of the HB process, the feed-in ratio automatically decreases to maintain the continuous operation of the HB process. Therefore, in terms of the overall cost, Case 4 achieves a lower overall cost than Case 3 by 14.6%, as illustrated in Fig. 6. Based on the insights gleaned from Fig. 6 and Fig. 8, it becomes evident that the proposed state-behavior mapping strategy presents notable advantages, including the prevention of costly shutdowns and the enhancement of electricity-hydrogen-ammonia coupling.

B. Comparative Analysis on the Effect of Proposed RAJIT Operation Scheme

In this subsection, we conduct a comparative analysis of seven methods to evaluate the out-of-sample performance of the proposed data-driven RAJIT scheme, as tabulated in Table IV. We benchmark its performance against state-of-the-art optimization methods, including deterministic optimization, robust optimization, moment-DRO, Wasserstein-DRO, and OL-DRO. Methods 1-5 utilize various state-of-the-art optimization methods. Method 6 and Method 7 employ the proposed data-driven RAJIT with different data-driven DRO methods.

Table IV Method Description

| Method No. | Method |
|---|---|
| 1 | Deterministic optimization |
| 2 | Robust optimization |
| 3 | Moment-DRO [a] |
| 4 | Wasserstein-DRO |
| 5 | OL-DRO |
| 6 | RAJIT+Wasserstein-DRO |
| 7 | **RAJIT+OL-DRO** |

a. Using the first- and second-order moment information of the uncertainty data

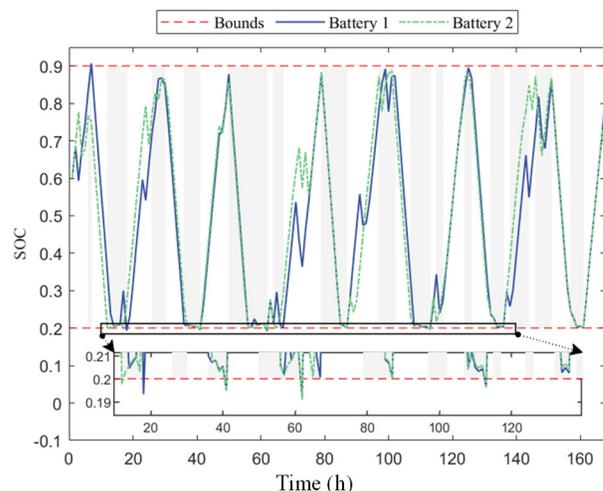

Fig. 9. State trajectories in Method 7. The grey area indicates periods with *flag*=1.

The battery SOC trajectories under the proposed method are depicted in Fig. 9. To facilitate a comprehensive evaluation, we provide a zoom-in view near the lower bound from the 12-*th* hour to the 138-*th* hour, encompassing six periods with low battery SOCs in unsafe regions. In these regions, the data-driven RAJIT switches from the risk-agnostic deterministic optimization to the risk-averse OL-DRO, aiming to reduce the constraint violation percentage. As it is clear from the zoom-in view, the two trajectories exceed the lower bound when they approach the boundary in the first three periods. In the subsequent three periods, the proposed method successfully learns the distribution of uncertainty from data streams, thus effectively avoiding constraint violations.

Table V demonstrates the computational results of Methods 1-7. The robust optimization method is the most conservative among all methods with a highest overall cost of $13,159.0. The overall cost of the deterministic optimization is the lowest, but this comes at the expense of the highest constraint violation percentage of 23.8%, which significantly exceeds the predefined threshold. Therefore, deterministic optimization is unsuitable for AHMG operation under uncertainty due to safety concerns. In contrast, the constraint violation percentages of other uncertainty-aware methods (Methods 2-7) remain within the acceptable threshold (15% at the confidence level of 85%), demonstrating their ability to ensure operational safety. Among these, the proposed method achieves the lowest overall cost while maintaining safety, substantially outperforming other uncertainty-aware optimization methods. These results clearly demonstrate that the proposed method is the least conservative and the most cost-effective among all uncertainty-aware optimization approaches.

## B. Sensitivity Analysis

To validate the adaptability of the proposed scheme, its economic performance and computational efficiency are further examined across different confidence levels and hysteresis bandwidths.

### 1) Confidence Level

The confidence level is tuned from 95% to 80% every 5%. TABLE V shows how the overall cost and constraint violation percentage change with confidence levels. Methods 3-7 are included because these DRO methods involve confidence levels. As can be observed in TABLE V, constraint violation percentages under Methods 3-7 exhibit certain rise as confidence level decreases, though they remain well below the threshold. It is worth noting that the proposed method outperforms other methods in overall cost under all confidence levels. Furthermore, adjusting the confidence level enables a finer trade-off between safety and economic performance, tailored to specific operational needs.

### 2) Hysteresis Bandwidth

We analyze the effect of tuning the hysteresis bandwidth from 0.2 to 0.05 in decrements of 0.05 on the operational dynamics and cost efficiency of the AHMG, as illustrated in Fig. 10. It reveals that a larger bandwidth results in the HB process spending a longer duration in the maximum feed ratio (Mode 1). This occurs because a higher bandwidth increases the threshold for transitioning from Mode 1 to Mode 2, thereby reducing the system's flexibility. The extended time in Mode 1, while stabilizing the HB process in this operational state, leads to an incremental increase in the overall cost.

TABLE V. COMPUTATIONAL RESULTS UNDER DIFFERENT CONFIDENCE LEVELS

| Confidence Level | Method No. | Overall Cost ($) | Constraint violation percentage (%) [a] |
|---|---|---|---|
|  | 1 | 13,070.2 | 23.8 |
|  | 2 | 13,159.0 | 0.0 |
| 0.95 | 3 | 13,157.1 | 1.94 |
|  | 4 | 13,130.9 | 2.35 |
|  | 5 | 13,126.2 | 4.05 |
|  | 6 | 13,117.9 | 3.42 |
|  | 7 | 13,095.8 | 4.41 |
| 0.90 | 3 | 13,137.0 | 2.74 |
|  | 4 | 13,122.5 | 2.97 |
|  | 5 | 13,118.9 | 6.54 |
|  | 6 | 13,113.2 | 4.48 |
|  | 7 | 13,091.6 | 6.69 |
| 0.85 | 3 | 13,134.0 | 4.73 |
|  | 4 | 13,121.0 | 7.66 |
|  | 5 | 13,106.0 | 8.80 |
|  | 6 | 13,103.1 | 8.20 |
|  | 7 | 13,089.3 | 12.14 |
| 0.80 | 3 | 13,132.2 | 4.78 |
|  | 4 | 13,118.2 | 10.07 |
|  | 5 | 13,087.4 | 13.17 |
|  | 6 | 13,083.7 | 10.37 |
|  | 7 | 13,071.1 | 14.04 |

a. Green numbers indicate values below the predefined threshold, while red numbers represent values exceeding the threshold.

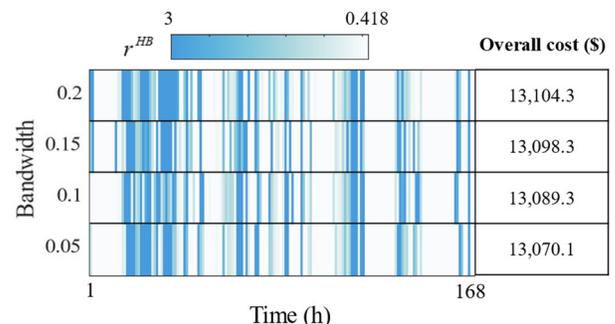

Fig. 10. Feed-in ratio $r^{HB}$ and overall cost under different hysteresis bandwidths.

## C. The Modified IEEE 69-Bus System

In this subsection, a modified IEEE 69-bus system integrated with three AHPs and eight battery systems is implemented to further verify the scalability of the proposed operation scheme, as depicted in Fig. 11. The IEEE 69-bus system is a well-defined test power system, and the detailed parameters can be found in the MATPOWER case file 'case69'. AHPs are located at Buses 8 and 54. Battery systems are located at Buses 5, 11, and 24. The comparative methods are the same with Table IV, and the confidence level is set as 85%.

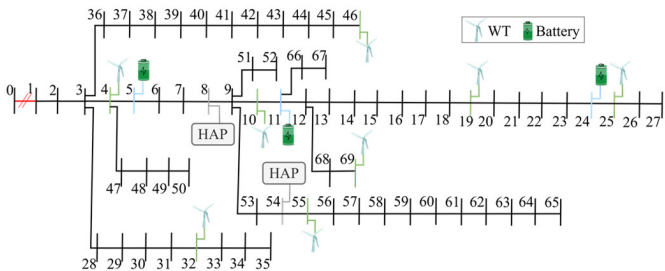

Fig. 11. Topology of the IEEE 69-bus distribution system with off-grid AHMG.

The computational results for the IEEE 69-bus system are tabulated in TABLE VI. Method 1 employs cost-oriented deterministic optimization throughout the operation horizon, achieving the lowest overall cost but resulting in the most severe constraint violation percentage of 18.9%, which indicates operational unsafety. In comparison, Methods 6 and 7, which incorporate the RAJIT scheme, demonstrate superior cost-effectiveness over other one-size-fits-all uncertainty-aware methods (Methods 2-5). Notably, Method 7, utilizing the data-driven RAJIT approach, achieves the lowest overall cost among all methods, thereby demonstrating the effectiveness of the proposed method. In terms of computational time, Cases 1-3 exhibit lower maximum single-step computational times than Case 4 because they enjoy fewer constraints. Among all methods, only Method 1 exhibits lower maximum single-step computational times than Case 9. This is due to the fact that the constraints in Method 1 are relatively simple. Nevertheless, the maximum and mean single-step computational times of the proposed data-driven RAJIT are 213.3 s and 26.1 s, respectively. This is acceptable within an optimization period of one hour or less. Together, the results underscore the practical applicability and computational efficiency of the proposed data-driven RAJIT approach, confirming its feasibility in real-world scenarios involving larger-scale off-grid AHMGs.

TABLE VI. COMPUTATIONAL RESULTS OF DIFFERENT OPTIMIZATION METHODS ON THE IEEE 69-BUS SYSTEM

| Case No. | Method No. | Overall Cost ($) | Constraint violation percentage (%) | CPU time (s) [a] |
|---|---|---|---|---|
| 1 | 7 | 23,300.5 | 8.4 | 214.7 |
| 2 | 7 | 23,308.9 | 10.2 | 210.2 |
| 3 | 7 | 29,532.5 | 10.4 | 211.4 |
| 4 | 1 | 22,125.7 | 18.9 [b] | 202.5 |
| | 2 | 22,805.2 | 0.0 | 228.1 |
| | 3 | 22,728.4 | 2.4 | 251.8 |
| | 4 | 22,711.9 | 6.4 | 342.2 |
| | 5 | 22,570.8 | 10.5 | 276.7 |
| | 6 | 22,632.8 | 9.3 | 293.4 |
| | 7 | 22,503.9 | 10.5 | 213.3 |

a. Maximum single-step computational time over the rolling horizon.
b. Note that the rolling optimization naturally suppress the constraint violation.

## V. CONCLUSION

This paper presented a novel RAJIT operation scheme for off-grid AHMGs to effectively enhance the electricity-hydrogen-ammonia coupling under uncertainty. First, we developed a practical and flexible model for off-grid AHMGs including a DGT and a state-behavior mapping strategy. Building upon this model, we proposed a RAJIT operation scheme to ensure safe and cost-effective real-time operation. With the proposed scheme, the off-grid AHMG can be risk-averse just in time according to its actual risk profile. To address the resulting intractable MINLFP, we derived an equivalent-reformulation-based solution methodology. Simulation tests were conducted on both IEEE 33-bus and IEEE 69-bus systems, validating the economic and safety advantages of the proposed scheme. While this study primarily focuses on off-grid scenarios, preliminary analysis indicates that the proposed method is also applicable to grid-connected conditions. In such scenarios, the main grid can provide backup power, allowing for the removal of the gas turbine. Our findings suggest that the proposed approach maintains its advantages in terms of cost-effectiveness and operational reliability, even when with grid support. Future work will explore the detailed performance and optimization strategies for grid-connected ammonia production systems.

REFERENCES

[1] C. Gu, Y. Liu, J. Wang, Q. Li, and L. Wu, "Carbon-Oriented Planning of Distributed Generation and Energy Storage Assets in Power Distribution Network With Hydrogen-Based Microgrids," *IEEE Trans. Sustain. Energy,* vol. 14, no. 2, pp. 790-802, 2023.
[2] Z. Dong, J. Yang, L. Yu, R. Daiyan, and R. Amal, "A green hydrogen credit framework for international green hydrogen trading towards a carbon neutral future," *Int. J. Hydrogen Energy,* vol. 47, no. 2, pp. 728-734, 2022.
[3] A. M. Abomazid, N. A. El-Taweel, and H. E. Z. Farag, "Optimal Energy Management of Hydrogen Energy Facility Using Integrated Battery Energy Storage and Solar Photovoltaic Systems," *IEEE Trans. Sustain. Energy,* vol. 13, no. 3, pp. 1457-1468, 2022.
[4] D. Carroll, "Western Australia to host green hydrogen project powered by 5.2 GW of wind, PV," *PV Magazine*, 4 May, 2022.
[5] R. H. C. Yushan Lou, Anne-Sophie Corbeau, Zhiyuan Fan. "Why China's Renewable Ammonia Market Is Poised for Significant Growth," 2024. https://www.energypolicy.columbia.edu/why-chinas-renewable-ammonia-market-is-poised-for-significant-growth/.
[6] "EDF Group, J-Power and Yamna Consortium Awarded a 1 Mtpa Green Ammonia Project in Oman," 2024. https://www.yamna-co.com/edf-group-j-power-and-yamna-consortium-awarded-a-1-mtpa-green-ammonia-project-in-oman/.
[7] "Europe's largest green ammonia plant," 2022. https://stateofgreen.com/en/solutions/europes-largest-green-ammonia-plant/.
[8] X. Sun, X. Cao, B. Zeng, Q. Zhai, and X. Guan, "Multistage dynamic planning of integrated hydrogen-electrical microgrids under multiscale uncertainties," *IEEE Trans. Smart Grid,* vol. 14, no. 5, pp. 3482-3498, 2022.
[9] G. Pan, W. Gu, Y. Lu, H. Qiu, S. Lu, and S. Yao, "Accurate Modeling of a Profit-Driven Power to Hydrogen and Methane Plant Toward Strategic Bidding Within Multi-Type Markets," *IEEE Trans. Smart Grid,* vol. 12, no. 1, pp. 338-349, 2021.
[10] K. Zhang, B. Zhou, C. Chung, S. Bu, Q. Wang, and N. Voropai, "A Coordinated Multi-Energy Trading Framework for Strategic Hydrogen Provider in Electricity and Hydrogen Markets," *IEEE Trans. Smart Grid,* vol. 14, no. 2, pp. 1403-1417, 2023.
[11] T. Wu, and J. Wang, "Reliability Evaluation for Integrated Electricity-Gas Systems Considering Hydrogen," *IEEE Trans. Sustain. Energy,* vol. 14, no. 2, pp. 920-934, 2023.
[12] K. Rouwenhorst, A. Van der Ham, G. Mul, and S. Kersten, "Islanded ammonia power systems: Technology review & conceptual process design," *Renewable & Sustainable Energy Reviews,* vol. 114, 2019.
[13] N. Campion, H. Nami, P. R. Swisher, P. V. Hendriksen, and M. Munster, "Techno-economic assessment of green ammonia production with different wind and solar potentials," *Renewable & Sustainable Energy Reviews,* vol. 173, 2023.
[14] D. Wen, and M. Aziz, "Flexible operation strategy of an integrated renewable multi-generation system for electricity, hydrogen, ammonia, and heating," *Energy Convers. Manage.,* vol. 253, pp. 115166, 2022.
[15] J. Li, J. Lin, P. M. Heuser, H. U. Heinrichs, J. Xiao, F. Liu, M. Robinius, Y. Song, and D. Stolten, "Co-Planning of Regional Wind Resources-based Ammonia Industry and the Electric Network: A Case Study of Inner Mongolia," *IEEE Trans. Power Syst.,* vol. 37, no. 1, pp. 65-80, 2021.